\documentclass[12pt]{article}

\usepackage{sbc-template}
\usepackage{graphicx,url}
\usepackage[utf8]{inputenc}
\usepackage[brazil]{babel}
\usepackage{braket}
\usepackage{qcircuit}
\usepackage{algorithm}
\usepackage{bussproofs}
\usepackage{algpseudocode}
\usepackage{mathtools}
\usepackage{numprint}
\usepackage{textcomp}
\usepackage{booktabs}
\usepackage{amsmath}
\usepackage{enumitem}
\usepackage{amssymb}

\DeclarePairedDelimiter\ceil{\lceil}{\rceil}
\DeclarePairedDelimiter\floor{\lfloor}{\rfloor}

\title{Quantum Algorithm for Multiplicative Linear Logic}

\author{Lorenzo Saraiva\inst{1}, Edward Hermann Haeusler\inst{1}, Vaston Costa \inst{2}}

\address{Pontificia Universidade Católica do Rio de Janeiro. \\ R. Marquês de São Vicente, 225 - Gávea, Rio de Janeiro - RJ, Brazil
\nextinstitute
  Universidade Federal de Catal\~ao. \\ Av. Dr. Lamartine Pinto de Avelar 1120. Catal\~ao - GO, Brazil
  \email{lorenzo.saraiva@hotmail.com, hermann@inf.puc-rio.br,
  vaston@ufcat.edu.br}
}

\begin{document} 

\maketitle

\begin{abstract}
  This paper describes a quantum algorithm for proof search in sequent calculus of a subset of Linear Logic using the Grover Search Algorithm. We briefly overview the Grover Search Algorithm and Linear Logic, show the detailed steps of the algorithm and then present the results obtained on quantum simulators.
\end{abstract}

\section{Introduction}

Quantum computing has provided us with algorithms that have a better time complexity than any classical counterpart, one of those being the Grover's Search Algorithm(GSA)\cite{grover1997quantum} for searching an element in an unordered database. The GSA is used in several contexts, including SAT, kmeans, genetic algorithms and pixel identification. 
% Linear logic is an extension of classical and intuitionistic logic that emphasizes the role of formulas as resources. For that reason, linear logic does not allow the rules of contraction and weakening to apply to all formulas but only those formulas marked with certain modals\cite{sep-logic-linear}.
In this work, we use the GSA to help in searching proofs of a subset of multiplicative linear logic to improve complexity compared to classic algorithms. We show the construction of the quantum circuit from the linear logic sequent to the end result and present our conclusions.

\section{Background} \label{sec:firstpage}

The GSA is one of the most famous quantum algorithms, and its goal is to search for an element in an unordered database. Assuming a database with $n$ qubits that contains $N = 2^n$ elements in the superposition, it has time complexity of $\sqrt{N}$, which outperforms any classical algorithm. The general steps of the Grover algorithm main iteration on $n$ qubits are as follows:

\begin{itemize}
    \item Database initialization - In this step an operator $A$ is applied to the database qubits to bring them from the initial state $\ket{0}^{\otimes n}$ to the desired state $\ket{\Psi}$. This state is usually the equal superposition state, and $A = H^{\otimes n}$. \newline
    %\item Ancilla initialization 
    \item Oracle call - In this step an oracle $O$ is applied to the prepared state $\ket{\Psi}$. The oracle will flip the phase of the searched value $x_t$ so that:\newline \newline
    $O \ket{x_s \neq x_t} = \ket{x_s}$\newline
    $O \ket{x_s = x_t} = \ket{-x_s}$ \newline
    \item Amplitude amplification - In this step an operator $D$ is used to amplify the amplitude of the state marked by the oracle. For such, an inversion about the mean (IAM) is performed. %First, we bring the state $\ket{\Psi}$ back to $\ket{0}^{\otimes n}$ applying $A^T$ and apply $D$.
\end{itemize}
Generalizing, the Grover iteration can be described as:
\[G = A O A^{T} D\]
The Grover iteration has to be repeated $\floor{\pi \sqrt{N}/4}$ times in order to maximize the probability of measuring the desired state. Our work follows Alsing's entangled database search \cite{alsing2011grover} using the GSA. The main feature of Alsing's algorithm is that, instead of using $A = H^{\otimes n}$ to prepare the equal superposition state, it chooses $A$ in order to encode an arbitrary list of pairs $\{s, t\}$. Thus, the algorithm's input is an entangled database with two sides, each side having one part of the pair. Every entry on the left side is entangled to an entry on the right side. In the GSA, it is necessary to know the searched value to construct the oracle. On Alsing's, on the other hand, one can construct the oracle based on a known entry $s_1$ on the left side, apply the GSA, and then measure the right side, recovering the unknown value $t_1$ entangled with $s_1$.

\section{Problem Description}

Linear logic is an extension of classical and intuitionistic logic that emphasizes the role of formulas as resources. For that reason, it does not allow the rules of contraction and weakening to apply to all formulas but only those formulas marked with special marks\cite{sep-logic-linear}. Due to the (formal) similarity between the logical rules that deal with these marks and the modalities in systems like S4, these marks might be considered as modalities. The absence of contraction and weakening allows Linear Logic to have two different versions of conjunction and disjunction: an additive and a multiplicative. The classical $\land$ (and), for example, is divided between the additive version, $\&$ (with), and the multiplicative version, $ \otimes$ (tensor). Linear logic also has a sequent calculus proof system. In this context, the algorithm for finding a cut-free proof in the multiplicative only version of Linear Logic has a worst-case time complexity of $2^k$, where $k$ is the number of atomic formulas. The subset of intuitionistic linear logic that deals only with the multiplicative connectives is called (intuitionistic) multiplicative linear logic(IMLL). In this work, we will be using a subset of IMLL, IMLL-$\otimes$ using only the tensor connector. 

% The rules for the tensor are depicted in the following circuit: %figure~\ref{fig:my_label}. 

\begin{figure}[H]
 \centering
 \includegraphics[width=10cm]{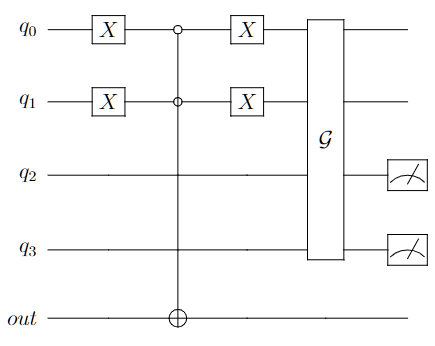}
\caption{Oracle circuit for k=2 and searching for the value associated with 0}
\label{fig:ite}
\end{figure}

% \end{center}
% \caption{Circuit for GSA iteration with the Alsing entangled database for finding the value associated with 00\label{fig:ite}}
% \end{figure}
% \end{center}

Considering a linear logic sequent with $k = 4$ atomic clauses

\[A \otimes (B \otimes (C \otimes D))  \vdash D \otimes (B \otimes (A \otimes C)).\]

We want to find the successive splits that verify that this is a valid proof. We have two rules that can be applied, $\otimes$-Left and $\otimes$-Right. In the classical algorithm, we apply the successive splits until we reach a valid axiom.

\begin{itemize}
    \item Apply one of the possible rules until there are only left axioms,
    \item If the axioms are all valid, the sequent is valid; if not, restart.
\end{itemize}
Since there are $2^k$ possible splits, the algorithm has time complexity of $O(2^k)$, where $k$ is the number of atomic clauses.

\begin{algorithm}
\caption{General Description}\label{alg:cap}
\begin{algorithmic}
\Require $k$ copies of the database of $2n$ qubits and $k$ pairs, where $n = \ceil{\log{k}}$
%\Ensure $y = x^n$
\State $numIterations \gets \floor{(\pi \times \sqrt{N}/4)}$
\For{$i < N$}
\For{$j < numIterations$}
\State buildOracle(n, target)
\State appendDiffuser($A$)
\State measure()

\EndFor
\EndFor
\end{algorithmic}
\end{algorithm}

% \begin{algorithm}
% \caption{Function buildOracle() for creating the Oracle}\label{alg:cap}
% \begin{algorithmic} 
% \Require n, target
% %\Ensure $y = x^n$
% \While{$i < n$}
% \If{$target[i] == 1$}
% \State XGate(circuit[i])
% \EndIf
% \EndWhile

% \State MCTGate(register, pk)

% \While{$i < n$}
% \If{$target[i] == 1$}
% \State XGate(circuit[i])
% \EndIf
% \EndWhile
% \end{algorithmic}
% \end{algorithm}

\section{Solution Steps}

Our quantum algorithm input is an entangled database with two sides, each part of a pair. Every entry on the left side is entangled to an entry on the right side, and they are both unordered. We will call left side of each pair as the \emph{search} part, or $s$, and the right side the \emph{target}, or $t$. To be able to perform the algorithm in $\sqrt{k}$ steps, it is necessary to have $k$ copies of the paired database, where $k$ is the number of unique atomic clauses. The complexity of building this database is not taken into account. Our algorithm also shows an explicit dynamic construction for the Grover oracle depending on the searched value.

\subsection{Preparing the entangled database}

Before starting the algorithm, one must construct an entangled database that accurately represents the sequent. In order to do so, we will need $k \times 2n$ qubits. Then, the pairs $\ket{a} \ket{b}$ will be encoded as $\ket{Na + b}$, where $a$ and $b$ is the position of the clause in each side of the sequent. Assuming we have $k = 8$ and $n = 3$, where $n$ is the number of qubits necessary to represent a solution space of $k$ values. Thus our entangled database with $8$ solutions will have $2$ groups of $n$ qubits, each representing $8 = k = 2^n$ values. We will treat both groups of $3$ qubits as a single array and prepare the resulting encoding in the superposition, using $k$ of the $k^2$ total possibilities that can be stored in $2n$ qubits. Then, we will need $k$ copies of the register, one for each clause. The construction of the database is not explicitly shown but its complexity is $O(n)$ or $O(\log{}k)$\cite{alsing2011grover}, taking $O(k\log{}k)$ in total. %Alsing proves in his paper that this "empty" space cannot be used to perform multiple concurrent Grover searches of several subsets of $N$ solutions\cite{alsing2011grover}.
This process is not strictly part of the algorithm, which only receives a pre-constructed entangled database.~\footnote{In our case, the entangled state, it is necessary to store the gate sequence $A$ used to encode a copy of the entangled database state so it can be used later in the IAM step of the GSA.} %With $A$ being the operator that takes the circuit from the $\ket{0}^{\otimes n}$ state to the initial state.

%\subsection{Entangled Database input}
%We will use $2^n = N$, so $n$ is the number of qubits necessary to represent a solution space of $N$ values. 2 qubits can represent a solution space of $2^n$, in this case 4. So our entangled database with 4 solutions will have 2 groups of 2 qubits. Each entry on the left side is associated randomly with and entry on the right side through entanglement.  It is relevant to note that this will only use a subspace of $N$ solutions of the $N^2$ possibles of the superposition. The construction of the database is not explicitly shown but its complexity is $O(n^2)$, where $n$ = $log (N)$. Alsing proves in his paper that this "empty" space can't be used to perform multiple concurrent grover searches of several subsets of $N$ solutions\cite{alsing2011grover}.

\subsection{$\otimes$-Left}
The first step of the algorithm itself is to apply the $\otimes$-Left rule until it cannot be applied anymore, so we have a sequent of the form: 

\[A^1, A^2, ..., A^N \vdash B \otimes  \Delta \]

Where $\Delta = A \otimes (\Delta)$ or $\Delta = A$ Now, we can use our entangled database to find out the correct split for the leftmost atomic clauses of the right side.
%Where $\Delta$ is a formula of the form $A \otimes (B \otimes ... C) ...) $ 

\subsection{Grover Search}

Now that we have the entangled database of $k$ copies of $2n$ qubits, we can perform the GSA. We start by picking the leftmost atomic clause of the right side. The first step is to construct the oracle dynamically for the chosen element on the left side. The construction of this Oracle takes into account the binary representation of the chosen clause position. We apply the necessary X-Gates to leave all the search space qubits in $\ket{1}$ and apply a multi-controlled Toffoli Gate with a prepared qubit as a target to perform the phase kickback, as can be seen in~\ref{alg:cap}. This oracle is applied only on the search, that is, the first $n$ qubits of the first copy of the $2n$ qubits. Then, the Grover operator for amplitude amplification is applied to all $2n$ qubits $\sqrt{k}$ times, and the measurement to the right side of the $2n$ qubits. 

An example of this circuit for $n = 2$ qubits is shown in \ref{fig:ite}. It is important to note that in the IAM step of the GSA, it is necessary to apply the $A$ operator, which takes $\log{}k$ steps, making the overall complexity of the GSA step $O(\sqrt{k}\log{}k)$. This process finds the corresponding entry of a pair, but we need to find the $k$ corresponding pairs. The issue is that measuring the qubits destroys the prepared superposition corresponding to the pairs. Therefore, we need the $k$ copies of the prepared $2n$ qubit entangled database - so we perform the GSA for each pair on a different copy of the database, in $k^{1.5}$ steps in total - $k$ times $\sqrt{k}$ steps.

\section{Example}

We want to find out if 
\[A \otimes (B \otimes (C \otimes D))  \vdash D \otimes (B \otimes (A \otimes C)))\]
is a valid sequent in linear logic. We have $k = 4$ and consequently $n = \log_2^4 = 2$, thus $2$ qubits are used for each side, and $2n$ for each copy in total. We will construct of the entangled representation of this sequent. 

\begin{center}
\begin{tabular}{ |c|c|c| } 
 \hline
 A & 0 & 2 \\ 
 B & 1 & 1 \\ 
 C & 2 & 3 \\ 
 D & 3 & 0 \\ 
 \hline
\end{tabular}
\end{center}

Using the formula $\ket{ka + b}$, with $k = 4$, we have \newline

\begin{center}
\begin{tabular}{ |c|c|c| } 
 \hline
 A & k0 + 2 = 2 \\ 
 B & k1 + 1 = 5\\ 
 C & k2 + 3 = 11 \\ 
 D & k3 + 0 = 12\\ 
 \hline
\end{tabular}
\end{center}
Thus we have the state of the quantum database as \newline

$\sqrt{\frac{1}{4}}(\ket{2} + \ket{5} + \ket{11} + \ket{12})$ \newline

%$\sqrt{\frac{1}{2}}(\ket{00} + \ket{11})$ \newline

or\newline

$\sqrt{\frac{1}{4}}(\ket{0010} + \ket{0101} + \ket{1011} + \ket{1100})$ \newline \newline
We perform the Grover search on any of the sides and are able to recover the value on the other side. But first, let's go back to the sequent. The rules for $\otimes$ are shown in \ref{fig:my_label}.

%\[
%\begin{array}{cc}
\begin{figure}
\begin{prooftree}
\AxiomC{$\Delta,B_0,B_1\vdash\gamma$}
\RightLabel{$\otimes$-Left}
\UnaryInfC{$\Delta,B_1\otimes B_2\vdash\gamma$}
\AxiomC{$\Delta_0\vdash A_0$}
\AxiomC{$\Delta_1\vdash B_1$}
\RightLabel{$\otimes$-Right}
\BinaryInfC{$\Delta_0,\Delta_1\vdash A_0\otimes A_1$}
\AxiomC{}
\UnaryInfC{$B\vdash B$}
\noLine
\TrinaryInfC{}
\end{prooftree}
%\end{array}
%\]
\caption{Rules of $\otimes$-only Linear Logic}
\label{fig:my_label}
\end{figure}

Because $\otimes$ is a binary operator, we can't search for values inside the parentheses and apply the rules, so we treat the sequent as\newline

$A \otimes  \Delta^1  \vdash D \otimes  \Delta^2$ \newline \newline

For that reason, we first apply the $\otimes$-Left successive times, until every atomic clause is alone \newline

\begin{prooftree}
\AxiomC{$A, B, C, D \vdash D \otimes  \Delta^2$}
\RightLabel{$\otimes$-Left}
\UnaryInfC{$A,  B, C \otimes D \vdash D \otimes  \Delta^2$}
\RightLabel{$\otimes$-Left}
\UnaryInfC{$A,  B \otimes \Delta^{3}  \vdash D \otimes  \Delta^2$}
\RightLabel{$\otimes$-Left}
\UnaryInfC{$A \otimes  \Delta^1  \vdash D \otimes  \Delta^2$}
\end{prooftree}

%$A \otimes  \Delta^1  \vdash D \otimes  \Delta^2$ \newline 

%$A,  B \otimes \Delta^{'}  \vdash D \otimes  \Delta^2$ \newline 

%$A,  B, C \otimes D \vdash D \otimes  \Delta^2$ \newline 

%$A,  B, C, D \vdash D \otimes  \Delta^2$ \newline 

Now, we run the quantum algorithm for every entry on the right side, and apply the results to the sequent, following the order of appearance. $D$ is encoded to the pair $(3, 0)$, but the algorithm only knows the $0$, which is the position of the value we're querying, which is encoded by $\sqrt{\frac{1}{4}}\ket{1100}$. We'll apply the Oracle on the two last qubits, that represent the $0$ part of the pair, with the shown circuit, and then measuring the first two qubits, with high probability of the result being $3$. This process is done for every clause of the right side, so now we just apply the splits following the indexes $(3, 1, 0, 2)$, with the following results\newline

\begin{prooftree}
\AxiomC{$A \vdash A$}
\AxiomC{$C \vdash C$}
\RightLabel{$\otimes$-Right}
\BinaryInfC{$A, C  \vdash A \otimes C$}
\AxiomC{$B \vdash B$}
\RightLabel{$\otimes$-Right}
\BinaryInfC{$A, B, C  \vdash B \otimes (A \otimes C)$}
\AxiomC{$D \vdash D$}
\RightLabel{$\otimes$-Right}
\BinaryInfC{$A,  B, C, D \vdash D \otimes  \Delta^2$}
\RightLabel{$\otimes$-Left}
\UnaryInfC{$A,  B, C \otimes D \vdash D \otimes  \Delta^2$}
\RightLabel{$\otimes$-Left}
\UnaryInfC{$A,  B \otimes \Delta^{3}  \vdash D \otimes  \Delta^2$}
\RightLabel{$\otimes$-Left}
\UnaryInfC{$A \otimes  \Delta^1  \vdash D \otimes  \Delta^2$}
\end{prooftree}

\section{Results}
From a given entangled database state, our algorithm has time complexity of $O(k^{1.5}\log{}k)$ and has a space qubit complexity of $\log{}k$, where $k$ is the number of atomic clauses on the sequent. Even when taking into account the construction of the database, which takes $O(k\log{}k)$ steps, we're still left with a time complexity of $O(k^{1.5}\log{}k + k \log{}k) = O(k^{1.5}\log{}k)$ which outperforms the classical algorithm. 

Additionally, it is important to note that when $k > 4$, we would need a controlled-NOT gate with more than two control qubits. For that, we need to concatenate the results of Toffoli gates, introducing additional $n - 2$ ancillary qubits\cite{piro2020generalizing}. We ran our circuit in the several simulators provided by IBM, such as the \emph{qasm\_simulator} and \emph{simulator\_mps}. Each execution consisted of 1000 circuit's runs. We tested the implementation of the algorithm up to 64 atomic clauses with high precision, using $2 \times \log{}k = 12$  qubits as search space.

\section{Future Work}
While this solution uses the GSA to get an advantage when searching the matching pairs, it has some weaknesses. The first is the fact that you need to prepare the quantum database for each execution, since the quantum state is destroyed in the process. Another issue is that the algorithm fully quantum: while the index matching is found with the GSA, the splits are done classically taking into account the position of each atomic clause, and one could argue that this could add an overhead of complexity. 
For that reason, a different quantum approach is being currently developed, where each qubit value will represent the side of an atomic clause in a specific split. 

\section{Full Quantum approach}
This algorithm also uses the GSA to help in proof search for IMLL, but there is considerable difference between this and the first one. Now, we don't use Alsing's entangled database nor do we need to prepare a specific quantum state prior to the execution. 
The algorithms uses $(k - 1) +\log{}k$ qubits, where $k$ is the number of atomic clauses in the right side of the sequent. The first $(k - 1)$ qubits represent the side picked by a clause in each of the $(k - 1)$ splits and the last $\log{}k$ qubits act as an index for the clauses. A qubit measured $0$ means a clause will go to the left in a split and $1$ means it will go the right. Starting with the simplest case:\newline

\[A, B  \vdash A \otimes B\]

For $k = 2$ We will need $(k - 1) + \log{}k = 2$ qubits. The quantum state that represents the correct splits is $\ket{00} + \ket{11}$. The $\ket{00}$ state is the $A$ going to the right side and the $\ket{11}$ is the $B$ going to the left side. The GSA Oracle will mark both these states as correct ones. These states are defined by the right side of the sequent. Let's go over a slightly more complicated example:

\[A, B, C, D  \vdash D \otimes (B \otimes (A \otimes C)))\]

We'll look at the right side to define the states that will be marked by the Oracle. First, $D$ will go to the left side and all everybody else to the right. $D$ will have no future splits, and in the case we fill the rest of its correspondent state with $0$s. Thus, one of the Oracle correct states is $\ket{010|11}$. Applying a similar process we can construct the other three: $\ket{110|00}$, $\ket{100|01}$ and $\ket{111|10}$, for A, B and C respectively. Now we just apply the GSA a sufficient time to measure the four possibilities and we'll have recovered the splits necessary to form a valid sequent. This has a time complexity of $\sqrt{\frac{2^{k + \log{}k}}{k}}$. This can be simplified:\newline \newline
$2^{k + \log{}k} = 2^k \times 2^{\log{}k} = 2^k \times k$ \newline\newline
$\sqrt{\frac{2^{k}\times k}{k}} = \sqrt{2^{k}}$\newline\newline
Which is the expected quadratic speedup from the GSA.

\section{Adding Linear Implication}

Following the full quantum approach, the next step is to add linear implication to the connectors accepted by the algorithm. This comes with some challenges. First, we can no longer use the right side as a fixed reference for the oracle to apply the successive splits based on the $\otimes$-Right rule - if we add linear implication, now the atomic clauses can switch sides depending on the rule, and the initial sequent no longer needs to have a balanced number of atomic clauses in each side. So, instead of only specifying the splits of the left side to follow a fixed order of the right side, we need to handle every atomic clause. Also, we have four options of "places to go" when applying the $\multimap$-Left rule: left side of the left sequent, right side of the left sequent, left side of the right sequent and right side of the right sequent. This is also an issue with the $\otimes$-Right, since now we have to explicitly say where each clause will go. To solve this, each step will use 2 qubits instead of one. The first qubit of the pair represents which sequent the clause will go, 0 for left, 1 for right. The second will represent which side of sequent the clause will go, again 0 for left, 1 for right. When applying the $\multimap$-Right, it will count as everybody going to the left sequent. Let's go over a simple example:

\[A^1, A^2 \multimap B^1 \vdash C^1 \multimap B^2, C^2\]
\[A^1, A^2 \multimap B^1, C^1 \vdash B^2, C^2\] 
\[A^1 \vdash A^2   \hspace{50pt}    B^1, C^1 \vdash B^2, C^2\] 

Thus, the correct states for the oracle will be: \newline
\[A^1 = \ket{0000|000}\]
\[A^2 = \ket{0001|001}\]
\[B^1 = \ket{0010|010}\]
\[B^2 = \ket{0111|011}\]
\[C^1 = \ket{0010|100}\]
\[C^2 = \ket{0111|101}\]

There's a few interesting things to point out here. The first is the increase of qubits. The complexity of the last solution was $\sqrt{2^{k}}$, where $k = n/2$, and $n$ is the total number of atomic clauses. This solution, on the other hand has complexity of $\sqrt{\frac{2^{c + \log{}n}}{n}}$. Simplifying on a similar way:\newline \newline
$2^{c + \log{}n} = 2^c \times 2^{\log{}n} = 2^c \times n$ \newline\newline 
$\sqrt{\frac{2^{c}\times c}{c}} = \sqrt{2^{2c}}$\newline \newline
$2^c$ is the final complexity.

\bibliographystyle{sbc}
\bibliography{sbc-template}

\end{document}